\documentclass[a4paper,twocolumn,pra,showpacs,showkeywords,aps,nofootinbib]{revtex4}
\usepackage{epsfig,amsmath,amsfonts,amssymb}
\usepackage{graphicx}
\usepackage[utf8]{inputenc}
\usepackage{epsfig,psfrag}
\usepackage{mathbbol}
 \usepackage{multirow}
\newcommand{\beq}{\begin{equation}}
\newcommand{\eeq}{\end{equation}}

\newcommand{\beqa}{\begin{eqnarray}}
\newcommand{\eeqa}{\end{eqnarray}}

\begin{document}
\title{A class of exactly solvable models to illustrate supersymmetry and test approximation schemes in quantum mechanics}
\author{C. M. Fabre}
\affiliation{Laboratoire de Collisions Agr\'egats R\'eactivit\'e,
CNRS UMR 5589, IRSAMC, Universit\'e Paul Sabatier, 118 Route de
Narbonne, 31062 Toulouse CEDEX 4, France}
\author{ D. Gu\'ery-Odelin}
\affiliation{Laboratoire de Collisions Agr\'egats R\'eactivit\'e,
CNRS UMR 5589, IRSAMC, Universit\'e Paul Sabatier, 118 Route de
Narbonne, 31062 Toulouse CEDEX 4, France}
 \date{\today}
\begin{abstract}
We derive the analytical eigenvalues and eigenstates of a family of potentials wells with exponential form (FPWEF). We provide a brief summary of the supersymmetry formalism applied to quantum mechanics and illustrate it by producing from the FPWEF another class of exact solutions made of their isospectral partners. Interestingly, a subset of the supersymmetric partners  provides a class of exactly solvable double well potentials. We use the exact solutions of the FPWEF to test the robustness and accuracy of different approximation schemes. We determine (i) the ground state through variational method applied to an approriate set of trial functions and (ii) the whole spectrum using three semiclassical quantization formula: the WKB, JWKB and its supersymmetric extension, the SWKB quantization formula. We comment on the importance of Maslov index and on the range of validity of these different semiclassical approaches.

\end{abstract}

\maketitle

Analytically solvable models in quantum mechanics are of pedagogical interest since they allow one to illustrate the abstract theoretical framework by concrete examples. For instance, the square well potential is very convenient for introducing the notion of reflection and transmission coefficients for the scattering states with a minimum of calculations, yielding some astonishing results such as the total quantum reflection of a low incident energy particle interacting with a potential well. In one dimension, besides the square potential, there are not many scattering potentials that are discussed in standard textbooks of quantum mechanics. Ref.~\cite{Landau} gives the interesting example of a step potential $U(x)=U_0/(1+e^{\alpha x})$ where one has to deal with subtle asymptotic conditions at infinity to solve the scattering problem.   

Interestingly, once an analytical solution is known, the supersymmetry (SUSY) techniques applied to quantum mechanics provide a whole family of analytical solutions having closely related properties.

Furthermore, analytically solvable models provide a testbed for comparing approximate methods to exact solutions, and can be used to model more complex situations.

In this article, we first investigate in one dimension a family of potential wells having an exponential form (FPWEF). In contrast with square well potentials, the potentials of the FPWEF are characterized by two parameters the depth of the potential $U_0$ and the typical length of variation of the potential $\alpha^{-1}$ (see Fig.~\ref{fig_uplus}). We solve analytically the scattering problem and determine the bound states of the potentials belonging to the FPWEF. This family provides an interesting example of the role of parity symmetry and the importance of boundary conditions on the existence and the number of bound states.

The second part of the article is devoted to the use of the supersymmetry formalism applied to quantum mechanics for the potential wells of the FPWEF \cite{Wit81,SUSYbook}. In this way, we find a new class of exactly solvable double well potentials.  

The third part explores the robustness and accuracy of approximated schemes in quantum mechanics. 
We compare the exact energy of the ground state for the potentials of the FPWEF with its approximate determination based on variational calculations.
We give explicit examples of the importance of the appropriate set of trial functions to get an accurate estimate of the ground state along with the limitations of this method. 
The exact bound spectrum of the potential wells is also compared with the predictions of different semiclassical quantization formula: 
the WKB, JWKB and the supersymmetric extension of the WKB formula (SWKB). We recover generic conclusions on the relative range of validity of these different approaches.

\section{The FPWEF}\label{HSEP}

\begin{figure}[t!]
    \begin{center}
        \includegraphics[width=9cm]{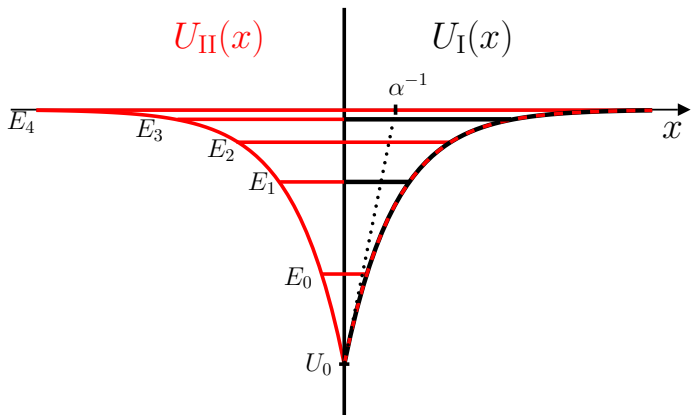}
    \end{center}
    \caption{Potentials $U_{\rm II}(x)$ and $U_{\rm I}(x)$ of depth $U_0$ and characteristic length $\alpha^{-1}$. Here, $U_{\rm II}(x)$ is plotted with 5 bound states ($a=8.48$) of energies $E_4>E_3>E_2>E_1>E_0$ (Red), and $U_{\rm I}(x)$ is represented for the same parameters (Black). This latter potential accomodates only two bound states of energies $E_3$ and $E_1$.}
    \label{fig_uplus}
\end{figure}

We first solve the quantum mechanical eigenstates for potentials $U_{\rm I}(x)$ of exponential form defined on $x>0$ with a sharp wall at $x=0$, and then for even potentials $U_{\rm II}(x)$ of exponential form defined on the whole real axis and with only soft walls: $U_{\rm I}(x\leq0)=\infty$ and $U_{\rm I}(x > 0)=-U_0e^{-\alpha x}$, and $U_{\rm II}(x)=-U_0e^{-\alpha |x|}$
with $\alpha >0$ and $U_0>0$ since we are considering potential wells.\footnote{The case $\alpha<0$ has been investigated from a mathematical point of view in \cite{Biane}.} The potential $U_{\rm I}(x)$ and $U_{\rm II}(x)$ are represented in Fig.~\ref{fig_uplus}. 

The full determination of the motion of a particle of mass $m$ that experiences $U_{\rm I\,or\,II}(x)$ requires the knowledge of the stationary states. These states are solution of the Schr\"odinger equation
\begin{equation}
\left[ -\frac{\hbar^2}{2m} \frac{{\rm d}^2}{{\rm d}x^2}+U_{\rm I \,or\, II}(x)
\right] \psi(x)=E\psi(x). \label{eqschropotentieluplus}
\end{equation}
Using the dimensionless variable $X=\alpha x$ and the dimensionless parameters $a=[8mU_0/(\hbar^2\alpha^2)]^{1/2}$ and $b=[8m(-E)/(\hbar^2\alpha^2)]^{1/2}$, we can rewrite Eq.~(\ref{eqschropotentieluplus}) in the form
\begin{equation}
\frac{{\rm d}^2\psi}{{\rm d}X^2}+\frac{1}{4}\left[a^2e^{-X}-b^2\right]\psi (X)=0.
\label{eqschropotentieluplus2}
\end{equation}
The solution of Eq.~(\ref{eqschropotentieluplus2}) can be expressed in terms of Bessel functions. Indeed, using the change of variable $y=a e^{-X/2}$, Eq.~(\ref{eqschropotentieluplus2}) takes the form of the second-order differential equation obeyed by the Bessel functions:
\begin{equation}
y^2\frac{{\rm d}^2\psi}{{\rm d}y^2}+ y\frac{{\rm d}\psi}{{\rm d}y}+[y^2-b^2]\psi (y)=0 \,.
\label{eqbessel}
\end{equation}

\subsection{The bound energies and states of $U_{\rm I}(x)$}

The energy of the bound states of the potential well $U_{\rm I}(x)$ are found in the negative domain of energy $E<0$ ($b$ real and positive). In this case, the solution of Eq.~(\ref{eqschropotentieluplus}) takes the general form
\begin{equation}
\psi(X\ge 0)=A^+_1J_{b}\left(ae^{-X/2}
\right)+A^+_2J_{-b}\left(ae^{-X/2} \right),
\label{eq6}
\end{equation}
where $J_b$ are the Bessel functions of the first kind.
The spectrum is determined by the boundary conditions:
$\psi(0)=0$, and $\psi(x\to \infty)=0$, from which we get the following set of equations
\begin{eqnarray}
&& A^+_1J_{b}(a)+A^+_2J_{-b}(a)=0, \label{cond1} \\
&& \left[A^+_1J_{b}(y \to 0)+A^+_2J_{-b}(y \to 0) \right]\to 0. \label{cond2}
\end{eqnarray}
The Bessel functions in the vicinity of zero scale as $J_\nu(y \to 0) \sim y^\nu$. As $b>0$, the divergence of $J_{-b}(y)$ when $y \to 0$ requires to set $A_2^+=0$ in order to fulfill the condition (\ref{cond2}). The discrete spectrum of energy $\{ E_n \}$ is therefore obtained from the condition (\ref{cond1}) and involves the zeros $\{ b_n \}$ of the Bessel function for a fixed value of the potential characteristics (depth $U_0$ and slope $\alpha$)
\begin{equation}
J_{b_n}(a)=0.
\label{besselzero}
\end{equation}
 The number of bound states is thus dictated by the value of the index $a$. The ground state energy is $E_0=-U_0b_0^2/a^2$. The wave function associated with the eigenenergy $E_n$ reads $\psi_n(x)= {\cal N}_nJ_{b_n}(a \exp(-\alpha x/2))$.
However, if $a<a_c\simeq 2.405$, there is no solution in $b$ for Eq.~(\ref{besselzero}) meaning that the potential $U_{\rm I}$ does not accomodate any bound state. Indeed, the sharp wall of the potential at $x=0$ rules out the application of the theorem according to which there is always at least a bound state for a one dimensional potential \cite{Messiah,Eza97}. The analysis of the bound states of $U_{\rm II}(x)$ (see below) enables a simple interpretation of the non-existence of a bound state for $U_{\rm I}(x)$ when $a<a_c$.

\subsection{The bound energies and states of $U_{\rm II}(x)$}

The general form of the solusion of the stationary Schr\"odinger equation for the potential $U_{\rm II}(x)$ is
\begin{eqnarray}
&& \psi(X\ge 0)=A^+_1J_{b}\left(ae^{-X/2}
\right)+A^+_2J_{-b}\left(ae^{-X/2} \right), \nonumber \\
&& \psi(X\le 0)=A^-_1J_{b}\left(ae^{X/2}
\right)+A^-_2J_{-b}\left(ae^{X/2} \right).
\label{solutu2}
\end{eqnarray}

The potential $U_{\rm II}(x)$ is even and thus commutes with the parity operator. As a result the eigenfunctions have a well-defined parity. The determination of the bound states is made by searching for solutions such that $\psi(x\to \pm \infty)=0$ which implies $A_2^+=A_2^-=0$, and the extra conditions $\psi(0)=0$ for the odd solutions, and $\psi'(0)=0$ for the even solutions. For a given potential (fixed value of  $a$), the corresponding discrete spectrum is given by the zeros $\{ b_n \}$ of the Bessel function (for the odd eigenfunction) and  the zeros $\{ \tilde{b}_n \}$ of its first derivative (for the even solution) 
\begin{equation}
J_{b_{n}}(a)=0,\qquad {\rm and}\qquad J'_{\tilde{b}_{n}}(a)=0,
\end{equation}
with $\tilde{b}_{0} > b_0 > \tilde{b}_{1} > b_1 >  \ldots$. The eigenstates are $\psi_n(x)= {\cal N}_nJ_{b_n}(a \exp(-\alpha |x|/2))$ for the eigenenergies $E_n=-U_0 b_n^2/a^2$ and $\tilde{\psi}_n(x)= \tilde{\cal N}_nJ_{\tilde{b}_n}(a \exp(-\alpha |x|/2))$ for the eigenenergies $\tilde{E}_n=-U_0 \tilde{b}_n^2/a^2$ where ${\cal N}_n$ and $\tilde{\cal N}_n$ are normalization factors. 
The subset of solutions $\{ b_n \}$ coincides with the eigenenergies of $U_{\rm I}(x)$ since they obey the same boundary conditions ($\psi(0)=0$ and $\psi(x\to +\infty)=0$). The extra subset $\{ \tilde{b}_n \}$ results from the extra symmetry of the potential $U_{\rm II}(x)=U_{\rm II}(-x)$ (see Fig.~\ref{fig_uplus}(a)). The ground state is given by the first root $\tilde{b}_{0}$ of the even solutions, i.e. $\tilde{E}_0= -U_0 \tilde{b}_0^2/a^2$. 

In contrast with the potential $U_{\rm I}(x)$ defined on half a space, there is always at least one bound state for the symmetric potential $U_{\rm II}(x)$ \cite{Messiah}. The threshold $a_c$ below which there is no bound state for $U_{\rm I}(x)$ can now be interpreted simply through the extended potential $U_{\rm II}(x)$. Indeed it corresponds to the threshold for the apparition of the first excited state of the potential $U_{\rm II}(x)$. The comparison between the spectrum of $U_{\rm II}(x)$ and $U_{\rm I}(x)$ therefore gives a pedagogical example on which one understands how boundary conditions influence the existence of at least one bound state in a one-dimensional potential. 

\subsection{Scattering states of $U_{\rm II}(x)$}

\begin{figure}[t!]
    \begin{center}
        \includegraphics[width=9cm]{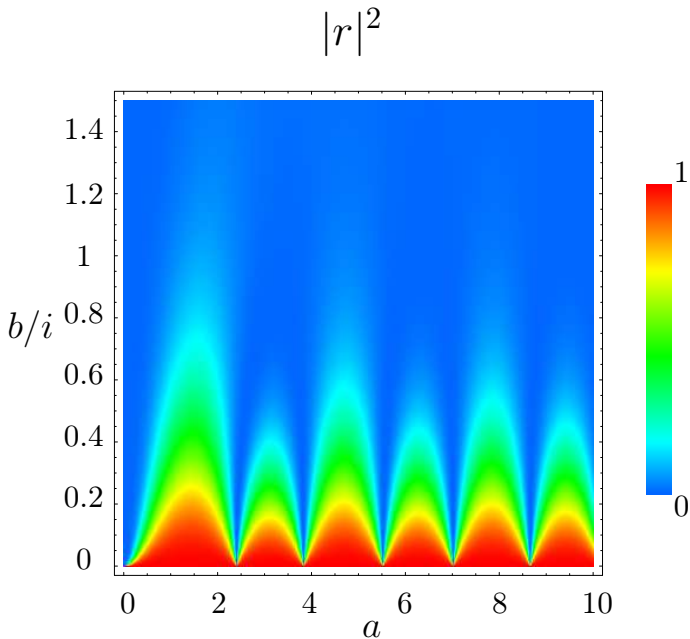}
    \end{center}
    \caption{(color online). Reflection probability $|r|^2$  for the scattering states of the potential $U_{\rm II}(x)$ as a function of the dimensionless parameters $a$ and $b$. The large reflection domains at low energy (low $\beta=b/i$) is a signature of quantum reflection. The periodic structure is a matter-wave Fabry-Perot like effect.}
    \label{fig_Rpuit}
\end{figure}

 The scattering states are obtained for $E>0$ ($b=i\beta$ is purely imaginary, $\beta>0$) and discussed here only for the potential $U_{\rm II}(x)$. An incident plane wave coming from $-\infty$ gives rise to a reflected and a transmitted waves. The asymptotic expansion for large $|x|$ values of the solution (\ref{solutu2}) yields
 \begin{eqnarray}
\psi(x) &\simeq& A^+_1\left(\frac{a}{2}\right)^{b}
\frac{e^{-ikx}}{\Gamma(1+b)}
+A^+_2\left(\frac{a}{2}\right)^{-b}\frac{e^{ikx}}{\Gamma(1-b)},
\nonumber \\
\psi(x)&\simeq& A^-_1\left(\frac{a}{2}\right)^{b}
\frac{e^{ikx}}{\Gamma(1+b)}
+A^-_2\left(\frac{a}{2}\right)^{-b}\frac{e^{-ikx}}{\Gamma(1-b)}.
\nonumber
\end{eqnarray}
We infer the reflection probability $|r|^2$ as a function of the dimensionless parameters $a$ and $b$ by setting $A_1^+=0$:
\begin{equation}
|r|^2= \left| \frac{A_2^-}{A_1^-} \right|^2=\frac{1}{4} \left|  \frac{J_b(a)}{J_{-b}(a)}  + \frac{J^\prime_b(a)}{J^\prime_{-b}(a)} \right|^2.
\label{rcoeffU2}
\end{equation}
The result, shown in Fig.~\ref{fig_Rpuit}, exhibits quantum reflection for low $\beta$ value. This is due to the fast variation of the de Broglie wavelength $\lambda_{\rm dB}(x)=h/mv(x)$ when $\beta$ tends to zero: 
\begin{equation}
{\rm Max}  \left( \frac{ {\rm d}\lambda_{\rm dB}  }{ {\rm d}x  } \right) = \frac{4\pi}{3\sqrt{3}} \frac{1}{\beta} \;\; \underset{\beta\to 0^+}{\longrightarrow}  \;\;+\infty.
\end{equation}
Furthermore, the reflection probability displays a periodic structure as a function of $a$ which is caused by Fabry-Perot cavity like resonance effect of the matter wave between the two``walls'' of the potential well.

\section{The supersymmetry formalism and its application to the FPWEF}\label{SUSY}

Supersymmetry (SUSY) applied to one-dimensional problems in quantum mechanics allows to construct a family of exactly solvable Hamiltonians from a given solvable problem \cite{SUSYbook}.
In this section, we provide a brief reminder on this method and then we illustrate it with the potentials of the FPWEF. 

\subsection{A short reminder}

 Let us consider a given solvable Hamiltonian
\begin{equation}
H = T+ V(x),\;\;{\rm where} \;\; T=-\frac{\hbar^2}{2m}\frac{ {\rm d}^2 }{ {\rm d} x^2} 
\end{equation}
is the kinetic term and $V(x)$ the potential energy term. The eigenvalues and eigenfunctions for the bound states are denoted $H|\psi_n\rangle = E_n|\psi_n\rangle$, with $n=0,1,2,\ldots$. Let us introduce the translated potential $V_-(x)=V(x)-E_0$. The corresponding Hamiltonian $H_-=T+V_-(x)$ has the same eigenfunctions, $|\psi_n^-\rangle=|\psi_n\rangle$ as $H$, and its eigenenergies, $E^-_n=E_n-E_0$, are translated with respect to those of $H$, and are therefore positive ($E^-_n \ge 0$):
\begin{equation}
H_-|\psi_n^-\rangle = E^-_n |\psi_n^-\rangle.
\label{spectremoins}
\end{equation}
The ground state of $H_-$ has a zero energy, $H_-|\psi_0^-\rangle=0$ so that
\begin{equation}
V_-(x)=\frac{\hbar^2}{2m} \frac{\psi_0^{\prime\prime}(x)}{\psi_0(x)},
\end{equation}
where $\psi_0^{\prime\prime}(x)$ is the second derivative of the ground state wave function, $\psi_0(x)$, with respect to the variable $x$. The hamiltonian $H_-$ can thus be recast in the simple form
\begin{equation}
H_-=-\frac{\hbar^2}{2m}\left( \frac{ {\rm d}^2 }{ {\rm d} x^2}  -  \frac{\psi_0^{\prime\prime}(x)}{\psi_0(x)} \right).
\end{equation}
In this form, this Hamiltonian can be somehow factorized, that is written as $H_-=A^+A^-$, where we have introduced the operators
\begin{equation}
A^\pm=-\frac{\hbar}{\sqrt{2m}} \left( \pm \frac{ {\rm d} }{ {\rm d} x}  +  \frac{\psi_0^{\prime}(x)}{\psi_0(x)} \right).
\end{equation}
This factorization can be viewed as a generalization of the one developed for the analysis of the one-dimensional harmonic oscillator in quantum mechanics \cite{InH51}.
Let us introduce the so-called superpotential \cite{Wit81}
\begin{equation}
W(x) = -\frac{\hbar}{\sqrt{2m}}  \frac{\psi_0^{\prime}(x)}{\psi_0(x)}. 
\label{superpot}
\end{equation}
This potential is defined over the domain ${\cal D}$ of values for which $V(x)$ remains finite. It has no divergence on this domain since the ground state wave function $\psi_0(x)$ has no nodes. The relation between the superpotential $W(x)$ and $V_-(x)$ is by definition
\begin{equation}
V_-(x) = W^2(x) - \frac{\hbar}{\sqrt{2m}} W^\prime(x).
\end{equation}
This relation suggests to introduce another potential defined by
\begin{equation}
V_+(x) = W^2(x) + \frac{\hbar}{\sqrt{2m}} W^\prime(x).
\end{equation}
The corresponding Hamiltonian $H_+=T+V_+(x)$ can also be simply expressed in terms of the operators $A^\pm$: $H_+ = A^-A^+$.
The spectrum (\ref{spectremoins}) of $H_-$ and of $H_+$ are closely related. Indeed, using the expression of $H_\pm$ in terms of the operators $A^\pm$, one readily shows that 
\begin{eqnarray}
&& H_-(A^+|\psi_n^+\rangle)=E_n^+(A^+|\psi_n^+\rangle), \label{shmoins} \\
&& H_+(A^-|\psi_n^-\rangle)=E_n^-(A^-|\psi_n^-\rangle). \label{shplus}
\end{eqnarray}
As $E_0^-=0$, we conclude that $A^-|\psi_n^-\rangle$ for $n \neq 0$ is an eigenstate of $H_+$ for the eigenvalues $E_n^-$. We can therefore write $|\psi_m^+\rangle=A^-|\psi_n^-\rangle$, so that $E_m^+=E_n^-$. Except for the ground state $E_0^-$, all eigenenergies of $H_-$ and $H_+$ coincide: $E_n^+=E_{n+1}^-$. Starting from a given solvable potential with $N_b$ bound states, one can thus construct by iteration a new set of $N_b$ exactly solvable potentials having respectively $N_b-1$, $N_b-2$, ..., $0$ bound states. 

Supersymmetry permits one also to relate the reflection and transmission coefficients when the two partner potentials, $V_\pm$, have continuous spectra. Let us assume for sake of simplicity that the potentials $V_\pm$ are defined over the whole real axis, and that the superpotential obeys the boundary conditions\footnote{One can readily generalized the reasonning performed here for finite but different asymptotic limits of the superpotential $W(x \to \pm \infty)=W_\pm< \infty$.} $W(x \to \pm \infty)=0$. It follows that $V_\pm(x \to \pm \infty)=0$. We consider an incident plane wave $e^{ikx}$ of energy $E=\hbar^2k^2/2m$ coming from the direction $x \to -\infty$. Asymptotically the scattering states that account for the reflected and transmitted waves read
\begin{eqnarray}
&& \psi^\pm(k,x\to -\infty) \sim e^{ikx}+ r^\pm(k)e^{-ikx}, \label{rpoum} \\
&& \psi^\pm(k,x\to +\infty) \sim t^\pm(k)e^{ikx}.  \label{tpoum} 
\end{eqnarray}
Using Eqs.~(\ref{shmoins}) and (\ref{shplus}) combined with Eqs.~(\ref{rpoum}) and (\ref{tpoum}), one finds $r_+(k)=-r_-(k)$ and $t_+(k)=t_-(k)$, which implies that the partner potentials have identical reflection and transmission probabilities ($|r_+(k)|^2=|r_-(k)|^2$ and $|t_+(k)|^2=|t_-(k)|^2$).

\subsection{Application to $U_{\rm I}(x)$}

\begin{figure}[t!]
    \begin{center}
        \includegraphics[width=7.5cm]{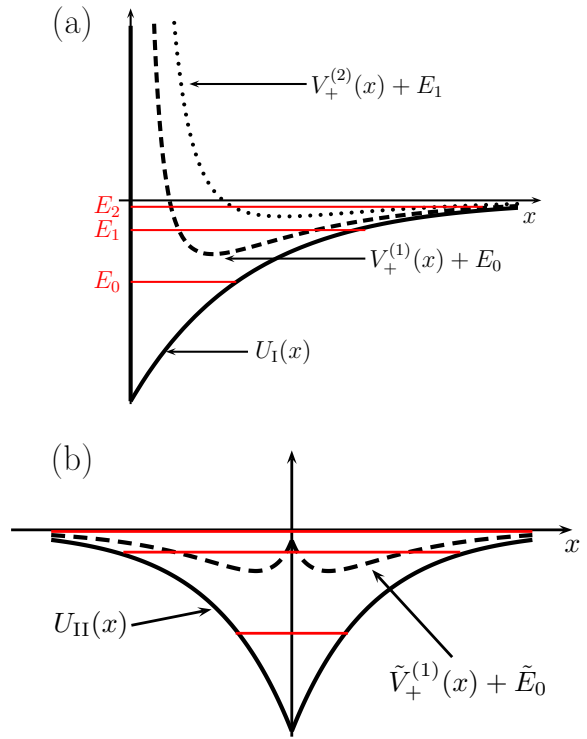}
    \end{center}
    \caption{(color online). (a) Potential $U_{\rm I}(x)$ with 3 bound states (red lines) and its first two supersymmetric partners $V^{(1)}_+(x)+E_0$ and $V^{(2)}_+(x)+E_1$ ($a=11.75$). (b) Potential $U_{\rm II}(x)$ with 3 bound states (red lines) and its first supersymmetric partner $\tilde{V}^{(1)}_+(x)+\tilde{E}_0$ that has a double well shape ($a=4.5$).}
    \label{susy}
\end{figure}

In order to use the formalism of supersymmetry, we introduce the potential $V_-^{(1)}(x)=U_{\rm I}(x)-E_0$ and deduce from Eq.~(\ref{superpot}) the expression for the corresponding superpotential $W^{(1)}$
\begin{equation}
W^{(1)}(x) = \sqrt{U_0} e^{-\alpha x/2} \frac{ J^\prime_{b_0}(ae^{-\alpha x/2})  }{ J_{b_0}(ae^{-\alpha x/2})  },
\end{equation}
and potential $V_+^{(1)}(x)=2(W^{(1)}(x))^2+E_0-U_{\rm I}(x)$. The potential $V_+^{(1)}(x)$ is a smooth potential that scales as $\sim x^{-2}$ when $x$ tends to zero and decays as $\exp (-\alpha x)$ for $\alpha x \gg 1$  (see Fig.~\ref{susy}(a)).\footnote{Note that this latter potential whose analytical solution is known can be used for solving the 3D scattering problem for the spherically symmetric potential $U_{\rm I}(r)$ in the $s$-wave regime since the reduced radial wave function $u(r)=r\psi(r)$ obeys the standard 1D Schr\"odinger equation.}

Similarly, one can introduce $V_-^{(2)}(x) = V_+^{(1)}(x) - (E_1-E_0)= 2(W^{(1)}(x))^2-U_{\rm I}(x)+2E_0-E_1$ whose ground state wave function is 
\begin{equation}
\psi_0^{(2)}  \propto A^-\psi_1 \propto\left( \psi_1^\prime - \frac{ \psi_0^\prime\psi_1}{ \psi_0} \right)
\end{equation}
from which we deduce the explicit form of the superpotential:
\begin{equation}
W^{(2)}=-\frac{\hbar}{\sqrt{2m}}  \frac{ \psi_0 (\psi_0 \psi_1^{\prime\prime}-\psi_0^{\prime\prime} \psi_1- \psi_0^{\prime} \psi_1^{\prime})+ \psi_0^{\prime} \psi_0^{\prime} \psi_1}{ \psi_0(\psi_0 \psi_1^{\prime}-\psi_0^{\prime} \psi_1)}.
\end{equation}
As previously, the potential $V_+^{(2)}(x)=2(W^{(2)}(x))^2-V_-^{(2)}(x)$ has the same spectrum as $V_-^{(2)}(x)$ except for the ground state. In Fig.~\ref{susy}(a), we represent $U_{\rm I}(x)$ for $a=11.75$, $\alpha=1$ that accomodates three bound states ($N_b=3$). We have also represented the supersymmetric partners $V_+^{(1)}(x)+E_0$ whose two bound states correspond to the first two excited states of $U_{\rm I}(x)$ and $V_+^{(2)}(x)+E_1$ whose unique bound state corresponds to the second excited state of $U_{\rm I}(x)$.

\subsection{Application to $U_{\rm II}(x)$}

The supersymmetric potential associated with $U(x)$ is directly deduced from the ground state wavefunction $\tilde{\psi}_0(x)$:
\begin{equation}
W^{(1)}(x) =  {\rm sgn}(x)\sqrt{U_0} e^{-\alpha |x|/2} \frac{ J^\prime_{\tilde{b}_0}(ae^{-\alpha |x|/2})  }{ J_{\tilde{b}_0}(ae^{-\alpha |x|/2})  }.
\end{equation}
Interestingly, we can derive here the whole supersymmetric family as in the previous example starting from a potential that has a singularity in its first derivative.\footnote{The SUSY literature contains another example  of potential singularities that can be handled: the Dirac singularities \cite{GLR94,UcT06}.} From Eq.~(\ref{superpot}), one indeed observes that if the potential has a differentiability class  ${\cal C}^n$ the supersymmetric potential has a differentiability class  ${\cal C}^{n+1}$. Repeating the same procedure, one readily derivates the family of supersymmetric potential partners $\{ \, \tilde{V}^{(n)}_+ \,\}$ of $U(x)$ (see an example on Fig.~\ref{susy}(b)). The supersymmetric partner $ \tilde{V}^{(1)}_+$
has a double well shape. The supersymmetry applied to the potential $U_{\rm II}(x)$ thus generates a family of exactly solvable weak double well potentials. Such a connection between single well and double well supersymmetric partners is discussed in \cite{BDG93}. There are not many examples of analytically solvable double well potentials. Let us mention for instance the potentials of the form $V(x)=k(|x|-a)^2$ \cite{Merzbacher}.

Note that the scattering reflection and transmission probabilities for the potential $\tilde{V}_+^{(1)}(x)$ are the same as those of the potential $U_{\rm II}(x)$. In particular, formula (\ref{rcoeffU2}) gives the reflection probability for any value of the trap parameters encompassed in the parameter $a$.

\section{Approximated techniques}

So far the results obtained are exact. In the following, we propose (i) to approximate the ground state energy of $U_{\rm I}(x)$ and $U_{\rm II}(x)$ using the variational method and to check and discuss the accuracy of this method and (ii) to test the accuracy of different semiclassical quantization formula for the whole spectrum.

\subsection{The variational method}

\begin{figure}[t!]
    \begin{center}
        \includegraphics[width=8cm]{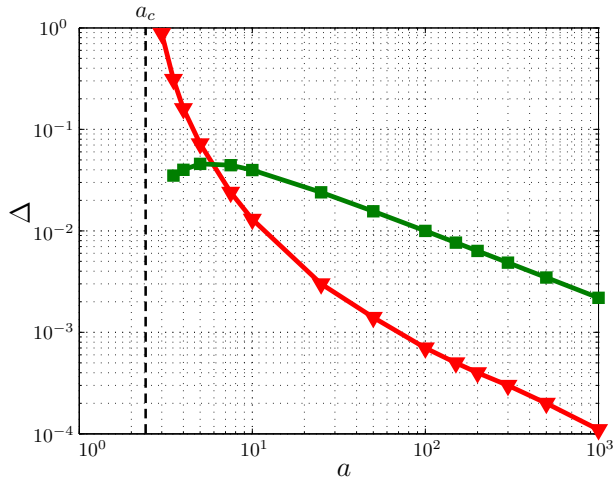}
    \end{center}
    \caption{(color online). Relative error, $\Delta=|E_{\rm exact}-E_{\rm I}(\eta_0)|/|E_{\rm exact}|$, on the estimate of the ground state energy using the Gaussian ansatz (\ref{ansatz1}) family $\{ \varphi_\sigma\}$ (red triangles), an exponential ansatz $\{ \bar{\varphi}_\sigma\}$ (green square) as a function of the dimensionless parameter $a$. Below the critical value $a_c\simeq 2.405$, the one-dimensional potential $U_{\rm I}(x)$ does not accomodate any bound states. At the crossing of the green (square) and red curve (triangle), the trap is still shallow and accomodates only two bound states.}
    \label{fig_upluserreur}
\end{figure}

In order to implement the variational method, we have to choose a set of trial wave functions $ \{\varphi_\sigma(x) \}$. We will consider a family of trial wave functions that depend on a unique parameter $\sigma$. The minimum of the expectation value of the hamiltonian $H=p^2/2m+U_{\rm I\, or \, II}(x)$ for these trial functions provides an upper bound of the ground state energy. 
The functional form of the trial ground state has to be chosen appropriately to get a good approximation of the ground state energy $E_0$:
$$
\underset{\sigma}{\rm Min} \left( \langle \varphi_\sigma | H | \varphi_\sigma\rangle \right) \geq E_0.
$$  

\subsubsection{Application to $U_{\rm I}(x)$}

As the potential has an infinite repulsive barrier at $x=0$, the ground state wave function vanishes at $x=0$. First, we choose the following family of functions:
\begin{equation}
\varphi_\sigma(x)= \left( \frac{2}{\pi} \right)^{1/4} \frac{x}{\sigma^{3/2}}e^{-x^2/4\sigma^2}
\label{ansatz1}
\end{equation}
which is normalized to unity and obeys the same boundary condition $\varphi_\sigma(0)=0$ and $\varphi_\sigma(+\infty)=0$ as the ground state. This guess is inspired by the first excited wave function of the one-dimensional harmonic oscillator. We now have to calculate the expectation value of $H$ for the wave functions $ \{\varphi_\sigma(x) \}$. This quantity is a function of $U_0$ and of the two dimensionless variables $a$ and $\eta=\alpha \sigma/\sqrt{2}$:
\begin{equation}
E _{\rm I}(\eta)= \langle   \varphi_\sigma | H| \varphi_\sigma \rangle=E_c (\eta)+E_p (\eta).
\end{equation}
One finds
\begin{eqnarray}
E_c (\eta) & = & \frac{\hbar^2}{2m} \int_0^\infty \left| \frac{ {\rm d} \varphi_\sigma}{ {\rm d} x} \right|^2 {\rm d}x = \frac{\hbar^2}{2m} \frac{3}{4 \sigma^2}=\frac{3U_0}{2a^2\eta^2},\nonumber \\
E_p (\eta) & = & -U_0 \left[  e^{\eta^2}(1+2\eta^2){\rm erfc}(\eta) - \frac{2\eta}{\sqrt{\pi}}\right]. 
\end{eqnarray}
The minimization of the total energy is obtained for $\eta_0$ which depends only on the dimensionless parameter $a$:
\begin{equation}
\frac{  {\rm d} E_{\rm I}}{  {\rm d} \eta} \bigg|_{\eta_0}=0,\;\; {\rm with} \;\; \frac{  {\rm d}^2 E_{\rm I}}{  {\rm d} \eta^2} \bigg|_{\eta_0}>0.
\label{eqsolvansat}
\end{equation}

Figure~\ref{fig_upluserreur} compares the relative error $\Delta=|E_{\rm exact}-E_{\rm I}(\eta_0)|/|E_{\rm exact}|$ in the estimate of the ground state energy as a function of the dimensionless parameter $a$ (red triangles). The poor accuracy for a small potential depth is due to the inappropriate functional form of the trial function that does not reproduce well the large extension of the wave function in the very shallow trap limit. This can be confirmed by using the family of normalized wave functions of the form $\bar{\varphi}_\sigma (x)=2xe^{-x/\sigma}/\sigma^{3/2}$ for the energy minimization. Such wave functions have a longer tail for large $x$ than those of the family $\{ \varphi_\sigma(x) \}$ and yield a better estimate in the low $a$ regime i.e. for a small trap depth as illustrated in Fig.~\ref{fig_upluserreur} (green squares). Furthermore, the variational method allows an approximate determination of the threshold value $a_c^{\rm ansatz}$ of the dimensionless parameter $a$ below which there is no  bound states. We find  $a_c^{\rm ansatz}\simeq 2.5142$ as the lowest bound of $a$ above which a solution of the equivalent of Eqs.~(\ref{eqsolvansat}) for the familiy $\{ \bar{\varphi}_\sigma(x) \}$ exists. This value differs by about $\sim$4.5 \% from the exact value. 
When the depth of the potential increases, the wave function becomes more localized and the estimate for the ground state energy is much better with the trial wave functions of the family $\{ \varphi_\sigma(x) \}$ as it clearly appears in Fig.~\ref{fig_upluserreur}. 

\subsubsection{Application to $U_{\rm II}(x)$}

The minimization of the hamiltonian expectation value for the potential $U_{\rm II}(x)$ is here performed in the sub-space of Gaussian trial functions 
\begin{equation}
\tilde{\varphi}_\sigma(x)= \left( \frac{1}{2\pi} \right)^{1/4} \frac{1}{\sigma^{1 /2}}e^{-x^2/4\sigma^2}.
\label{ansatz1}
\end{equation}
This family of trial functions is inspired 
by the ground state wave function of the one-dimensional 
harmonic oscillator and has no node as expected for the ground state of a potential well. We find, for example, $E(\eta_0)\simeq - 0.545 U_0$ for $a=5$ which differs from the exact value by about 1 \%.

The variational method also allows for the determination of first excitated state. For this purpose, one has to choose a family that has the same symmetry as the state considered and that is orthogonal to the family of states used for the determination of the ground state. From this respect, the extension of the trial functions used for the potential $U_{\rm I}(x)$ to the family $\{$ $\hat{\varphi}_\sigma(x \geq 0)=\varphi_\sigma(x)$ and $\hat{\varphi}_\sigma(x \leq 0)=-\varphi_\sigma(-x)$  $\}$ provides a new family that is precisely orthogonal to the family $\{$ $\tilde{\varphi}_\sigma$ $\}$, has one node and is thus appropriate for the determination of the first excited state through the variational method. This calculation is exactly the one performed on the half space $x\geq 0$ for the potential $U_{\rm I}(x)$ whose accuracy is summarized on an example in Fig.~\ref{fig_upluserreur}. Thereby, the search for an approximate of the ground state energy of the potential $U_{\rm I}(x)$ through the variational principle gives an estimate of the first excited state of the symmetric extension $U_{\rm II}(x)$ of the potential $U_{\rm I}(x)$.

\subsection{The semiclassical quantization formula}

\begin{figure}[t!]
    \begin{center}
        \includegraphics[width=7cm]{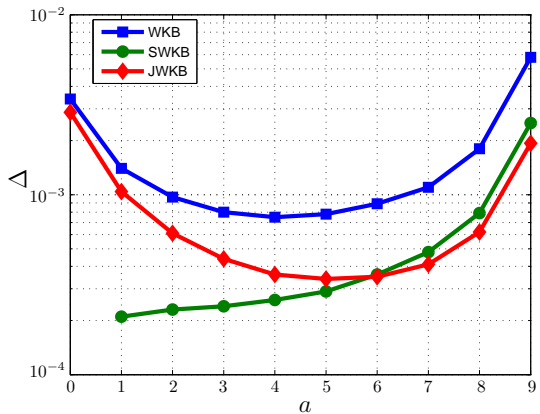}
    \end{center}
    \caption{(color online). Comparison of WKB (blue squares), SWKB (green disks) and JWKB (red diamonds) predictions for the energy spectrum of $U_{\rm I}(x)$ with $a=32$ (10 bound states) with exact results. The relative error $\Delta$ is plotted for the 10 bound states as a function of the dimensionless parameter $a$.}
    \label{figtable1}
\end{figure}

In order to get an approximate determination of the whole spectrum, one relies on semiclassical quantization formula. In this section, we test the accuracy of three semiclassical approaches.

The most commonly used is the WKB quantization condition which reads\footnote{For a 1D harmonic potential or a Morse potential, this approximation gives a spectrum that coincides with the exact one.} \cite{Wen26,Kra26,Bri26,Jef25}
\begin{equation}
\oint p\,{\rm d}x = \left( n+\nu \right) h,
\label{semiclq}
\end{equation}
where $\nu$ is the Maslov index, whose value is determined by the matching conditions of the wave function at the turning points. A smooth (sharp) wall gives a contribution 1/4 (1/2). For the potential $U_{\rm I}(x)$, we have a sharp wall at $x=0$ and a smooth one for $x>0$ so that $\nu=1/4+1/2=3/4$. The calculation of the action $\oint p\,{\rm d}x$ in the case of the potential $U_{\rm I}(x)$ combined with Eq.~(\ref{semiclq}) gives the following implicit equation for the approximate semiclassical determination of the eigenenergies:
\begin{equation}
 \left( n+\frac{3}{4} \right) \frac{\pi}{a} = F(y_n),
 \label{WKBeq1}
\end{equation}
with $F(y)=\sqrt{1-y^2} - y\cos^{-1}(y)$, $y_n=\exp(-\alpha x_n/2)$ and $U(x_n)=E$. 
Semiclassical approaches are supposed to work better for large quantum numbers. We compare in Fig.~\ref{figtable1} the exact energies of the potential $U_{\rm I}(x)$ with a choice of parameters ($a=32$ and $\alpha=1$) such that it accomodates 10 bound states with the approximated values obtained from Eq.~(\ref{WKBeq1}). We indeed observe an accuracy which gets better and better up to the fifth level but then which get worse. 

The standard WKB estimate for the energies is significantly improved by taking into account higher orders corrections in $\hbar$ to the standard WKB quantization condition, this approximation scheme is referred to as the  JWKB quantization condition \cite{KiL77}. The first correction, $\delta$, reads
\begin{equation}
\left(n+\nu \right)=\frac{1}{h}\oint p\,{\rm d}x+\delta
\end{equation}
where
\begin{equation}
\delta=-\frac{1}{24\pi}\left(\frac{\hbar^2}{2m}\right)^{1/2}\frac{\partial}{\partial E}\left( \int_{x_1}^{x_2}\frac{U''(x)}{\left(E-U(x)\right)^{1/2}}{\rm d}x\right).\nonumber
\end{equation}

The explicit form for the potential $U_{\rm I}(x)$ reads: 
\begin{equation}
\left(n+\frac{3}{4}\right)=F(y_n)-\frac{1}{12\pi a \sqrt{1-y_n}}.
\end{equation}
Compared to the WKB results, we obtain an improved accuracy for the whole spectrum (see Fig.~\ref{figtable1}).

Combining the supersymmetry formalism with the WKB method, one can work out a SWKB quantization condition \cite{CBC85,Eck86}. This third semiclassical quantization formula reads
\begin{equation}
\int_{x_{\rm min}}^{x_{\rm max}}  \left[ 2m(E^-_n-W^2(x))\right]^{1/2} {\rm d}x = n\hbar \pi 
\label{nwkbf}
\end{equation}
where $E^-_n=W^2(x_{\rm min})=W^2(x_{\rm max})$. 
The SWKB approach yields even the exact bound state spectra for all shape invariant potentials (SIPs), that is when the pair of the SUSY partners $V_+$ and $V_-$ are similar in shape and differ only in the parameters \cite{DKS86,YCS10}. The potential $U_{\rm I}(x)$ does not belong to the SIPs and thus provides an interesting example to test the accuracy of the SWKB spectrum prediction. By construction, the SWKB approach requires the knowledge of the ground state wave function and thus gives the exact ground state energy. As a result it provides the best estimate for the deep energy states as it clearly appears in Fig.~\ref{figtable1}. Among the three semiclassical approximation schemes, the JWKB formula turns out to be the most accurate for the states near the continuum. These conclusions on the relative range of validity and accuracy of the different semiclassical quantization formula is quite generic.

Let us consider the highest bound state for a deeper potential.\footnote{
In 3D quantum scattering theory, this state plays a particular role since it determines the magnitude and sign of the scattering length, a result also referred to as the Levinson's theorem \cite{GrF93}.}
For the three approximation schemes (WKB, JWKB and SWKB), the deeper the last bound state the better the estimate. The JWKB quantization formula gives systematically a better account of the energy of the last state. This result is well known in molecular physics. Note that the JWKB quantization condition for the highest vibrational levels in a molecular potential can be further improved using the Gribakin and Flambaum formula  \cite{GrF93} for the scattering length\cite{BAV98,BAV00}.
 
  The WKB quantization rule for the potential $U_{\rm II}(x)$ yields
\begin{equation}
 \left( n+\frac{1}{2} \right) \frac{\pi}{2a} = F(y_n),
\end{equation}
with $y_n=\exp(-\alpha x_n/2)$ and $U(x_n)$ is the energy of the $n^{\rm th}$ state since the potential well $U_{\rm II}(x)$ has two smooth walls ($\nu=1/4+1/4=1/2$). 
As expected, the odd values of $n$ coincide with the energies determined by applying the WKB quantization condition with $U_{\rm I}(x)$ (see Eq.~(\ref{WKBeq1})). 
This example provides a clear illustration of the importance of the Maslov index. An extra subset of energies is obtained that corresponds to the even values of $n$, it includes the ground state $n=0$ which has an energy lower than the ground state $U_{\rm I}(x)$ (see Fig.~\ref{fig_uplus}).

\section{Discussion}

We have shown the application of supersymmetry to the potentials $U_{\rm I}(x)$ and $U_{\rm II}(x)$ that were not investigated so far in this field, and illustrated the importance of exact solutions to test approximate methods in quantum mechanics. The same approach can be used to analyze the family of potentials of the form $|x|$. In addition to the pedagogical value of these examples, we would like to emphasize that supersymmetry complements perfectly the traditional teaching of quantum mechanics at the undergraduate physics. It answers precisely to basic and important questions such as: can two potential wells have the same spectrum and different shape ? Is it possible to rebuild the potential shape knowing its reflection and transmission probabilities for all incident energies ? Does it exist transparent potential ? It generalizes the factorization procedure of the stationary Schr\"odinger equation introduced for the harmonic oscillator, enlarges the class of exactly solvable potentials and provides new approximating schemes for quantization rules. 

\begin{acknowledgments}
It is a pleasure to thank P. Labastie, J. Vigu\'{e} and A. Comtet for fruitful discussions. 
We are grateful to R. Mathevet, T. Lahaye, O. Carraz, and P. Cheiney for useful comments.
We acknowledge financial support from the R\'{e}gion Midi-Pyr\'{e}n\'{e}es, the C.N.R.S., the Agence Nationale de la Recherche (ANR-09-BLAN-0134-01) and Institut Universitaire de France.
\end{acknowledgments}

\end{document}